\documentclass[12pt]{article}
\usepackage{amsfonts}

 \def\be{\begin{equation}}
 \def\ee{\end{equation}}
 \def\bea{\begin{eqnarray}}
 \def\eea{\end{eqnarray}}

\def\1{\'{\i}}                           
\def\>#1{{\bf #1}}                 
\def\d{{\rm d}}
\def\la{\lambda}

\def\te{\theta}
\def\rr{r}
\def\tes{\phi}

\def\jp{J_+}
\def\jm{J_-}
\def\jj{J_3}

\def\ff{g}

 \def\cte{\alpha}
\def\cteb{\beta}

\def\otra{b}

\newcommand{\dd}{{\mathrm d}}
\newcommand{\Kep}{_{\mathrm{KC}}}
\newcommand{\Harm}{_{\mathrm{O}}}

\begin{document}

\begin{center}

{\sc{\Large{Superintegrable potentials on 3D Riemannian and }}}

{\sc{\Large{Lorentzian spaces with non-constant curvature}}}

\end{center}

\medskip

\begin{center}
{\'Angel Ballesteros$^a$, Alberto Enciso$^b$, Francisco J. Herranz$^c$
and Orlando Ragnisco$^d$}
\end{center}

\small
\noindent
{$^a$ Depto.~de F\'\i sica, Facultad de Ciencias, Universidad de Burgos,
09001 Burgos, Spain\\ ~~E-mail: angelb@ubu.es\\[10pt]
$^b$ Depto.~de F{\'\i}sica Te\'orica II,   Universidad Complutense,   28040 Madrid,
Spain\\ ~~E-mail: aenciso@fis.ucm.es\\[10pt]
$^c$ Depto.~de F\'\i sica,  Escuela Polit\'ecnica Superior, Universidad de Burgos,
09001 Burgos, Spain \\ ~~E-mail: fjherranz@ubu.es\\[10pt]
$^d$ Dipartimento di Fisica,   Universit\`a di Roma Tre and Instituto Nazionale di
Fisica Nucleare sezione di Roma Tre,  Via Vasca Navale 84,  00146 Roma, Italy  \\
~~E-mail: ragnisco@fis.uniroma3.it

\normalsize

\begin{abstract}
A quantum $sl(2,\mathbb R)$ coalgebra (with deformation parameter $z$)  is shown to
underly the construction of a large class of superintegrable potentials on 3D curved spaces, that include the    non-constant  curvature analogues  of the spherical, hyperbolic and
(anti-)de Sitter spaces. The connection and curvature tensors for these ``deformed" spaces are fully studied by working on two different phase spaces. The former directly  comes from a 3D symplectic realization of the deformed coalgebra, while the latter is obtained through a map leading to a spherical-type phase space. In this framework, the non-deformed limit $z\to 0$ is identified with
the flat contraction leading to the     Euclidean and Minkowskian spaces/potentials. 
 The resulting   Hamiltonians  always admit, at least,  three  functionally independent  constants of   motion coming from the
coalgebra structure. Furthermore, the intrinsic oscillator and Kepler potentials on such Riemannian and Lorentzian spaces of non-constant  curvature are identified, and several examples of them are explicitly presented.
\end{abstract}



 \vfill
 \eject
 
%%%%%%%%%%%%%%%%%%%%%%%%%%%%%%%%%%%%%%%%%%%%%%%%%%%%%%%%%%%%%%%%%%%%%%%%%%%%%%%%%

\section{Introduction}

In the context of Hamiltonian systems with an arbitrary finite number of degrees of freedom, a deep connection between the coalgebra symmetry of a given system and its Liouville integrability was firmly established in~\cite{coalgebra}. Moreover, the intrinsic superintegrability properties of the coalgebra construction were further explored in~\cite{CRMAngel}. Since then, this framework has lead to the coalgebra interpretation of the integrability properties of many well-known systems, as well as to the construction of many new superintegrable systems by using both Lie and $q$-Poisson coalgebras (see~\cite{coalgebra, CRMAngel, chains, BH07} and references therein).

In particular, by making use of the Poisson coalgebra given by the non-standard quantum deformation of $sl(2,\mathbb R)$, the construction of integrable 2D geodesic flows corresponding to 2D Riemannian and Lorentzian spaces with non-constant curvature was presented in~\cite{plb}. Furthermore, these systems revealed a geometric interpretation of the quantum deformation, since the (in general, non-constant) curvature of these spaces was just a smooth function of the deformation parameter. Later, integrable potentials on such 2D  ``quantum deformed" spaces were introduced by preserving the underlying deformed coalgebra symmetry~\cite{jpa2D}.  In this context, the search for the appropriate Kepler--Coulomb  (KC) and oscillator potentials on such 2D curved spaces was posed as an interesting problem, and some hints were proposed.

In this contribution we present the generalization of all these results to the 3D case and we fully solve the question concerning the generic form of the intrinsic oscillator and KC potentials on all the corresponding ``quantum deformed" 3D spaces, thus opening the path for the generalization of this construction to $N$ dimensions. In the next section we make use of the quantum  $sl(2,\mathbb R)$ coalgebra symmetry in order to construct the family of superintegrable 3D geodesic flows that define the spaces of hyperbolic type, whose sectional and scalar curvatures are also obtained. In section 3 an analytic continuation procedure is introduced through a set of appropriate spherical-type coordinates, thus leading to the Lorentzian counterparts of the previous spaces. Moreover, it is shown that these coordinates allows the separability of the geodesic flow Hamiltonians in all the cases. Finally, section 5 is devoted to the characterization of those potentials that will preserve the superintegrability properties of the free Hamiltonian. In particular, by applying  the prescription given in~\cite{LT87,LT95,EP06d,CQG}, the intrinsic oscillator and KC potentials on all the previous spaces are explicitly constructed, and some particular examples are   analysed.

%%%%%%%%%%%%%%%%%%%%%%%%%%%%%%%%%%%%%%%%%%%%%%%%%%%%%%%%%%%%%%%%%%%%%%%%%%%%%%%%%

\section{Superintegrable Hamiltonians}

Let us   consider the Poisson coalgebra version of the non-standard
quantum deformation $sl(2,\mathbb R)$, hereafter denoted $(sl_z(2,\mathbb R),\Delta)\equiv sl_z(2)$, where $z$ is a {\em real} deformation parameter
($q={\rm e}^z$). Its deformed  Poisson brackets,  coproduct $\Delta$ and  Casimir ${\cal
C}$  are   given by~\cite{chains}:
\begin{equation}
 \{J_3,J_+\}=2 J_+ \cosh z J_-  , \quad  
\{J_3,J_-\}=-2\,\frac {\sinh zJ_-}{z} ,\quad   
\{J_-,J_+\}=4 J_3   , 
\label{ba}
\end{equation}
\begin{equation}
\begin{array}{l}
\Delta(J_-)=  J_- \otimes 1+ 1\otimes J_- ,\quad 
\Delta(J_l)=J_l \otimes {\rm e}^{z J_-} + {\rm e}^{-z J_-} \otimes
J_l   ,\quad l=+,3,
\end{array}
\label{bb}
\end{equation}
\begin{equation} 
{\cal C}= \frac {\sinh zJ_-}{z} \,J_+ -J_3^2  . 
\label{bc}
\end{equation} 
A  one-particle symplectic realization of (\ref{ba}) reads
\begin{equation}
 J_-^{(1)}=q_1^2 ,\quad   J_+^{(1)}=
  \frac {\sinh z q_1^2}{z q_1^2}\,   p_1^2+\frac{z \otra_1}{\sinh
z q_1^2}  ,\quad J_3^{(1)}=
\frac {\sinh z q_1^2}{z q_1^2 }\,    q_1 p_1  ,
\label{bd}
\end{equation}
where $\otra_1$ is a real parameter that labels the representation through ${\cal
C}^{(1)}=\otra_1$. Hence     dimensions of the deformation parameter
are $[z]=[J_-]^{-1}=[q_1]^{-2}$.

Starting from (\ref{bd}),  the coproduct (\ref{bb}) determines the
corresponding two-particle  realization of (\ref{ba}) defined on
$sl_z(2)\otimes sl_z(2)$  that depends on two real parameters
$\otra_1$, $\otra_2$:
$$ 
\begin{array}{l}
\displaystyle{ \jm^{(2)}=q_1^2+q_2^2 ,\qquad \jj^{(2)}=
\frac {\sinh z q_1^2}{z q_1^2 } \, q_1 p_1  \, {\rm e}^{z q_2^2} +
\frac {\sinh z q_2^2}{z q_2^2 }\,  q_2 p_2  \, {\rm e}^{-z q_1^2} } ,\\[10pt]
\displaystyle{  \jp^{(2)}=
\left( \frac {\sinh z q_1^2}{z q_1^2}\,  p_1^2 +\frac{z \otra_1}{\sinh
z q_1^2} \right) {\rm e}^{z q_2^2} +
\left(\frac {\sinh z q_2^2}{z q_2^2} \, p_2^2  +\frac{z \otra_2}{\sinh z
q_2^2} \right) {\rm e}^{-z q_1^2} } .
\end{array}
\label{bbe}
$$
Then the two-particle Casimir  is given by
\bea
&& {\cal C}^{(2)} = \left(  \frac {\sinh z q_1^2
}{z q_1^2 } \,
\frac {\sinh z q_2^2}{z q_2^2} 
\left({q_1}{p_2} - {q_2}{p_1}\right)^2 +\otra_1\, \frac {\sinh z q_2^2}{\sinh z q_1^2}
+ \otra_2\, \frac {\sinh z q_1^2}{\sinh z q_2^2}  \right)
{\rm e}^{-z q_1^2}{\rm e}^{z q_2^2} \cr 
&&\qquad\qquad  +\otra_1 {\rm e}^{2z q_2^2} +\otra_2 {\rm e}^{-2z
q_1^2}    . 
\label{bbf}
\eea 
Next, the 3-sites
coproduct, $\Delta^{(3)} =(\Delta \otimes \mbox{id})\circ
\Delta =(\mbox{id}\otimes \Delta )\circ \Delta $, 
gives rise to  a three-particle symplectic realization  of (\ref{ba}) in terms of three real parameters
$\otra_i$ which is  defined
on $sl_z(2)\otimes sl_z(2)\otimes sl_z(2)$~\cite{coalgebra,chains}; namely
\be  
\begin{array}{l}
\!\! \jm^{(3)}=q_1^2+q_2^2+q_3^2\equiv \>q^2 , \\[1pt] 
\!\!\displaystyle{  \jj^{(3)}=
\frac {\sinh z q_1^2}{z q_1^2 }  q_1 p_1   {\rm e}^{z q_2^2} {\rm
e}^{z q_3^2}+
\frac {\sinh z q_2^2}{z q_2^2 }  q_2 p_2   {\rm e}^{-z q_1^2}{\rm
e}^{z q_3^2} +
\frac {\sinh z q_3^2}{z q_3^2 }  q_3 p_3   {\rm e}^{-z q_1^2}{\rm
e}^{-z q_2^2}  },\\[8pt]
\!\!\displaystyle{ \jp^{(3)}= 
\left( \frac {\sinh z q_1^2}{z q_1^2}\,  p_1^2 +\frac{z \otra_1}{\sinh
z q_1^2} \right) {\rm e}^{z q_2^2}{\rm e}^{z
q_3^2}  +
\left(\frac {\sinh z q_2^2}{z q_2^2} \, p_2^2  +\frac{z \otra_2}{\sinh z
q_2^2} \right) {\rm e}^{-z q_1^2} }{\rm e}^{z
q_3^2} \\[8pt]
\!\!\displaystyle{\qquad\qquad + \left(\frac {\sinh z q_3^2}{z q_3^2} \, p_3^2  +\frac{z
\otra_3}{\sinh z q_3^2} \right) {\rm e}^{-z q_1^2} }{\rm e}^{-z
q_2^2}  .
\end{array}
\label{be}
\ee
 
Hence
if we  denote the three sites on
$sl_z(2)\otimes sl_z(2)\otimes sl_z(2)$ by $1\otimes 2\otimes 3$ the coalgebra
approach~\cite{coalgebra, annals} provides {\em three} ``relevant" functions, coming from the two-
and  three-sites coproduct of the Casimir (\ref{bc}):
(i)~the two-particle Casimir ${\cal
C}^{(2)}$  which is defined on $1\otimes 2$; (ii)~another two-particle Casimir
${\cal C}_{(2)}$  but defined on $2\otimes 3$; and (iii)~the   three-particle Casimir
${\cal C}^{(3)}$  defined   on $1\otimes 2\otimes 3$. These are given by (\ref{bbf}) and
\bea
&&\!\!\!\!  {\cal C}_{(2)} =\left( \frac {\sinh z q_2^2
}{z q_2^2 } \,
\frac {\sinh z q_3^2}{z q_3^2} 
\left({q_2}{p_3} - {q_3}{p_2}\right)^2   + \otra_2\, \frac {\sinh z q_3^2}{\sinh z q_2^2}
+ \otra_3\, \frac {\sinh z q_2^2}{\sinh z q_3^2}   \right)
{\rm e}^{-z q_2^2}{\rm e}^{z q_3^2}  \cr 
&&\qquad\qquad +\otra_2 {\rm e}^{2z q_3^2} +\otra_3 {\rm e}^{-2z
q_2^2} , 
\nonumber\\
&&
\!\!\!\! {\cal C}^{(3)}   = \left(  \frac {\sinh z q_1^2
}{z q_1^2 } \,
\frac {\sinh z q_2^2}{z q_2^2} 
\left({q_1}{p_2} - {q_2}{p_1}\right)^2 +\otra_1\, \frac {\sinh z q_2^2}{\sinh z q_1^2}
+ \otra_2\, \frac {\sinh z q_1^2}{\sinh z q_2^2}  \right)  {\rm e}^{-z q_1^2}{\rm e}^{z
q_2^2} {\rm e}^{2 z q_3^2} \nonumber \\  
&&\qquad +
\left( \frac {\sinh z q_1^2 }{z q_1^2 } \,
\frac {\sinh z q_3^2}{z q_3^2} 
\left({q_1}{p_3} - {q_3}{p_1}\right)^2  +\otra_1\, \frac {\sinh z q_3^2}{\sinh z q_1^2}
+ \otra_3\, \frac {\sinh z q_1^2}{\sinh z q_3^2}  \right) {\rm e}^{-z q_1^2} {\rm e}^{  z
q_3^2}\nonumber  \\ &&\qquad +
\left( \frac {\sinh z q_2^2
}{z q_2^2 } \,
\frac {\sinh z q_3^2}{z q_3^2} 
\left({q_2}{p_3} - {q_3}{p_2}\right)^2   + \otra_2\, \frac {\sinh z q_3^2}{\sinh z q_2^2}
+ \otra_3\, \frac {\sinh z q_2^2}{\sinh z q_3^2}   \right) {\rm e}^{-2z q_1^2}{\rm e}^{-z
q_2^2} {\rm e}^{z q_3^2}  \nonumber \\
&&\qquad +  \otra_1 {\rm e}^{2z q_2^2} {\rm e}^{2z q_3^2} +\otra_2 {\rm e}^{-2z
q_1^2} {\rm e}^{2z q_3^2}  +\otra_3 {\rm e}^{-2z
q_1^2}{\rm e}^{-2z
q_2^2} .
\label{bbg}
\eea

All the above expressions give rise to the   (non-deformed) $sl(2,\mathbb R)$ coalgebra~\cite{BH07}  under
the limit $z\to 0$, that is, the Poisson brackets and Casimir are non-deformed,
the coproduct is primitive, $\Delta(X)=X\otimes 1 + 1\otimes X$, and  the
symplectic realization reads  $J_-^{(3)}=\>q^2$,  $J_+^{(3)}=  \>p^2+\sum_{i=1}^3
\otra_i/q_i^2$, $J_3^{(3)}=\>q\cdot\>p$. If we set all  the  $\otra_i=0$, the three
Casimir functions reduce to the   components of the angular momentum $ { l}_{ij}=q_ip_j-q_jp_i$: ${\cal
C}^{(2)}= { l}^2_{12}$, ${\cal C}_{(2)}= { l}^2_{23}$ and ${\cal
C}^{(3)}={ l}^2_{12}+{ l}^2_{13}+ { l}^2_{23}$.

In this way a large family of {\em superintegrable} Hamiltonians  can be
constructed through the following statement.
 
 \noindent
{\bf Proposition 1.}
{\em 
(i) The three-particle generators (\ref{be}) fulfil the  commutation rules (\ref{ba})
with respect to the   canonical Poisson bracket $\{q_i,p_j\}=\delta_{ij}$.\\
(ii) These generators Poisson commute with the three   functions ${\cal C}^{(2)}$, ${\cal
C}_{(2)}$ and ${\cal C}^{(3)}$.\\ 
(iii) Any     function defined on (\ref{be}), i.e., 
\be
{\cal H}= {\cal H}(\jm^{(3)},\jp^{(3)},\jj^{(3)}),
\label{ham}
\ee
provides a completely integrable Hamiltonian since either  $\{{\cal C}^{(2)},{\cal
C}^{(3)},{\cal H} \}$ or $\{{\cal C}_{(2)},{\cal
C}^{(3)},{\cal H} \}$ are three functionally
independent functions in involution.\\
(iv) The   four functions $\{{\cal C}^{(2)},{\cal C}_{(2)},{\cal
C}^{(3)},{\cal H} \}$  are functionally
independent.}

We remark that in the 2D case, the generic  function ${\cal H}= {\cal
H}(\jm^{(2)},\jp^{(2)},\jj^{(2)})$ determines, in principle, an {\em integrable}
(but not superintegrable!) Hamiltonian as it  is only endowed with  a single constant of the motion ${\cal
C}^{(2)}$~\cite{jpa2D}. On  the contrary, in  the 3D case any Hamiltonian $\cal H$ (\ref{ham})  is (at least)
 a {\em weakly-superintegrable} one~\cite{Evansa}, since one more constant
of the motion is lacking in order to ensure the maximal superintegrability of the system. It is well-known~\cite{coalgebra, CRMAngel} that in the $N$D generic case the coalgebra approach would provide
$2N-2$ constants of the motion, thus leading to the construction of {\em quasi-maximally superintegrable} systems.  Obviously, in 3D, quasi-maximal superintegrability is equivalent to weak superintegrability, but this is no longer true in higher dimensions. Therefore the 3D case can be considered as the cornerstone for the generalization of all the results we shall present here to arbitrary dimension $N$.

%%%%%%%%%%%%%%%%%%%%%%%%%%%%%%%%%%%%%%%%%%%%%%%%%%%%%%%%%%%%%%%%%%%%%%%%%%%%%%%%%

\section{Curved spaces from geodesic flows}

As a byproduct of proposition 1 we can obtain  an infinite family of  superintegrable
{\em free} Hamiltonians    by setting the three $\otra_i=0$ and by choosing, amongst the
family (\ref{ham}),  the following expression for ${\cal H}$:
\be  
\begin{array}{l} {\cal H}=\frac 12 \jp\, f (z\jm  ),
\end{array}
\label{ahaa}
\ee 
where $f$ is any  smooth function such that $\lim_{z\to
0}f(zJ_- )=1$; hence  $\lim_{z\to 0}{\cal H}=\frac
12 \>p^2$ gives the kinetic energy on the 3D Euclidean space. For the sake of simplicity, from now  on we drop the index ``${(3)}$" in the generators. Thus by writing the
Hamiltonian (\ref{ahaa})  as a free Lagrangian, 
$$
 {\cal H}=\frac 12 \left( \frac
 {z q_1^2}{\sinh z q_1^2} \, {\rm e}^{-z q_2^2}{\rm e}^{-z q_3^2} \dot
q_1^2   +
 \frac {z q_2^2}{\sinh z q_2^2} \, {\rm e}^{z q_1^2}{\rm e}^{-z q_3^2}
\dot q_2^2    +
 \frac {z q_3^2}{\sinh z q_3^2} \, {\rm e}^{z q_1^2}{\rm e}^{z q_2^2}
\dot q_3^2  \right) f (z\>q^2 ),
$$
 we find the geodesic flow on a 3D
   space with a definite positive  metric   given  by
\be
 \d s^2=\left( \frac {2z q_1^2}{\sinh z
 q_1^2} \, {\rm e}^{-z q_2^2}{\rm e}^{-z q_3^2} \,\d q_1^2   +
  \frac {2 z q_2^2}{\sinh z q_2^2} \, {\rm e}^{z q_1^2}{\rm e}^{-z
q_3^2}\, \d q_2^2   +
  \frac {2 z q_3^2}{\sinh z q_3^2} \, {\rm e}^{z q_1^2}{\rm e}^{z
q_2^2}\, \d q_3^2 \right) \frac{1}{f (z\>q^2 )}.
 \label{ccd}
\ee
The connection   $\Gamma^i_{jk}$ $(i,j,k=1,2,3)$ can now  be straightforwardly computed:
\bea
&& \Gamma^i_{ii}=\left(\frac{1}{q_i^2}-\frac{z}{\tanh
zq_i^2}-z\,\frac{f^\prime(x)}{f(x)}
\right)q_i,\qquad \Gamma^i_{jk}=0,\quad i,j,k\ {\rm different},\nonumber\\
&& \Gamma^i_{ij}=-z q_j\left(1+\frac{f^\prime(x)}{f(x)}
\right),\quad \Gamma^j_{ij}=z q_i\left(1-\frac{f^\prime(x)}{f(x)}
\right),\quad i<j,\nonumber\\
&& \Gamma^i_{jj}=-z\, {\rm e}^{z (q_i^2+q_j^2)}\,\frac{q_j^2\sinh z
q_i^2}{q_i\sinh z q_j^2}\left( 1-\frac{f^\prime(x)}{f(x)}  \right)
,\quad j=i+1,\nonumber\\
&& \Gamma^j_{ii}=z\, {\rm e}^{-z (q_i^2+q_j^2)}\,\frac{q_i^2\sinh z
q_j^2}{q_j\sinh z q_i^2}\left( 1+\frac{f^\prime(x)}{f(x)}  \right)
,\quad j=i+1,\nonumber\\
&& \Gamma^1_{33}=-z\, {\rm e}^{z \>q^2}{\rm e}^{z q_2^2}\,\frac{q_3^2\sinh z
q_1^2}{q_1\sinh z q_3^2}\left( 1-\frac{f^\prime(x)}{f(x)}  \right),\nonumber\\
&& \Gamma^3_{11}=z\, {\rm e}^{-z \>q^2}{\rm e}^{-z q_2^2}\,\frac{q_1^2\sinh z
q_3^2}{q_3\sinh z q_1^2}\left( 1+\frac{f^\prime(x)}{f(x)}  \right),\nonumber
\eea
where  $x\equiv z\jm^{(3)} =z  \>q^2$,     $f^\prime(x)=\frac{{\rm d}f(x)}{{\rm d}
 x}$ and $f^{\prime\prime}(x)=\frac{{\rm d}^2f(x)}{{\rm d} x^2}$.
Therefore, the three  geodesic
equations for $q_i(s)$ read,
$$
\frac{\d^2 q_i}{\d s^2}+\sum_{j,k=1}^3\Gamma_{jk}^i \,\frac{\d q_j}{\d s}\,\frac{\d q_k}{\d
s}=0 ,
\label{cm}
$$
  where $s$ is the canonical parameter of the metric (\ref{ccd}). Next the  
 Riemann   tensor,   the  sectional  curvatures $K_{ij}$
in the planes 12, 13 and 23, and the scalar curvature $K$ can be deduced~\cite{Doub}. The
latter turn out to be, in general, non-constant and read
\bea
&& K_{12}=\frac{z}{4}\,{\rm e}^{-z \>q^2}\left\{ \left(1+ {\rm e}^{2 z q_3^2}-2 {\rm e}^{2z
\>q^2} \right)  \left( f(x)+{f^\prime}^2(x)/f(x) \right) \right.\nonumber\\
&&\qquad\qquad +\left. 2 \left(  1+ {\rm e}^{ 2 z \>q^2}\right)f^\prime(x) -
 2 \left( {\rm e}^{2 z q_3^2} - {\rm e}^{ 2 z \>q^2} \right)f^{\prime\prime}(x)\right\}
,\nonumber\\
&& K_{13}=\frac{z}{4}\,{\rm e}^{-z \>q^2}\left\{ \left(2- {\rm e}^{2 z q_3^2}+ {\rm e}^{2 z
(q_2^2+q_3^2)}-2  {\rm e}^{2z
\>q^2} \right)  \left( f(x)+{f^\prime}^2(x)/f(x) \right) \right.\nonumber\\
&&\qquad\qquad +\left. 2 \left(  1+ {\rm e}^{ 2 z \>q^2}\right)f^\prime(x) -
2 \left(1- {\rm e}^{2 z q_3^2}+{\rm e}^{2 z
(q_2^2+q_3^2)} -  {\rm e}^{ 2 z \>q^2} \right)f^{\prime\prime}(x)\right\}
,\nonumber\\
&& K_{23}=\frac{z}{4}\,{\rm e}^{-z \>q^2}\left\{ \left(2- {\rm e}^{2 z (q_2^2+q_3^2)}- {\rm
e}^{2 z \>q^2 } \right)  \left( f(x)+{f^\prime}^2(x)/f(x) \right) \right.\nonumber\\
&&\qquad\qquad +\left. 2 \left(  1+ {\rm e}^{ 2 z \>q^2}\right)f^\prime(x) - 2
\left(1-  {\rm e}^{2 z
(q_2^2+q_3^2)} \right)f^{\prime\prime}(x)\right\}
,\nonumber
\eea
$$
 K(x)=z\left( 6 f^\prime(x)\cosh x  +\left(
  4  f^{\prime\prime}(x)-5f(x)-5{f^\prime}^2(x)/f(x)
 \right) \sinh x
 \right) .
\label{co}
$$
Notice that although the sectional curvatures are not symmetric with respect to the
coordinates $q_i$, the scalar curvature is actually a radial function, since it does only depend on $x=z\>q^2$
and fulfils $K=2(K_{12}+K_{13}+K_{23})$.

Therefore the coalgebra construction giving rise to the family of metrics (\ref{ccd})  can be
understood as the introduction of a variable curvature (controlled by the quantum deformation parameter $z$) on the formerly flat Euclidean space~\cite{plb,ober}.
This, in turn, means 
 that   the non-deformed
limit $z\to 0$  can then be identified with the {\em flat contraction} providing the proper
3D Euclidean space with metric $\d s^2=\sum_{i=1}^3 \d q_i^2$.  

Let us present two specific instances for
${\cal H}$  (\ref{ahaa}) which   have    been studied in~\cite{plb,ober}.

\noindent
$\bullet$ The simplest free Hamiltonian  arises by setting $f(x)\equiv 1$.  The  
scalar curvature  yields
\bea
&& K= -5 z \sinh(z\>q^2) ,
\nonumber
\eea 
which, for any non-zero value of $z$, determines a hyperbolic Riemannian space of non-constant negative curvature.
 
\noindent
$\bullet$ When $f(x)={\rm e}^{x}$ we find a very singular case as all the  curvatures are  
{\em constant}:   
$K_{ij}=z$ and $K=6z$.
It turns out that the resulting Hamiltonian is 
a St\"ackel system~\cite{Per} which provides an additional 
constant of motion~\cite{chains} (that cannot be derived from the underlying coalgebra symmetry):
$$
 {\cal I}=\frac {\sinh z q_1^2}{2 z q_1^2} \, {\rm e}^{z
q_1^2}  p_1^2  .
$$
Since this additional integral is functionally independent with
respect to the  three   constants of the motion given in proposition 1, the geodesic motion is {\em maximally
superintegrable}, as it  should be the case. This result encompasses the  three classical
Riemannian spaces of constant curvature, {\em i.e.}, the   spherical, hyperbolic and Euclidean spaces, provided that a positive, negative and zero quantum deformation parameter
 $z$ is  respectively considered. Notice that a fully equivalent situation arises when $f(x)={\rm e}^{-x}$ is chosen, thus leading to sectional curvatures given by $K_{ij}=-z$.

%%%%%%%%%%%%%%%%%%%%%%%%%%%%%%%%%%%%%%%%%%%%%%%%%%%%%%%%%%%%%%%%%%%%%%%%%%%%%%%%%

\section{Spherical-type coordinates and Lorentzian spaces}

It is also possible to obtain non-constant curved spaces of pseudo- and semi-Riemannian type
(with Lorentzian and degenerate metrics) through a graded contraction (or analytic continuation) approach~\cite{plb,ober}. Explicitly, let us   consider the  spherical-type coordinates
$(r,\te,\tes)$  defined by:
\bea  &&
\cos^2(\la_1r)= {\rm e}^{-2z \>q^2}, \cr &&
\tan^2(\la_1r)\cos^2(\la_2\te)={\rm e}^{2 z q_1^2}{\rm e}^{2 z
q_2^2}\bigl({\rm e}^{2 z q_3^2}-1 \bigr) ,\cr &&
\tan^2(\la_1r)\sin^2(\la_2\te)\cos^2\tes={\rm e}^{2 z q_1^2}
\bigl({\rm e}^{2 z q_2^2}-1 \bigr),\label{xc} \\ &&
\tan^2(\la_1r)\sin^2(\la_2\te)\sin^2\tes= {\rm e}^{2 z q_1^2}-1 ,
\nonumber
\eea 
where $z=\la_1^2$ and $\la_2$ is an additional parameter
which can  be either a real  or a pure imaginary number~\cite{plb}. Under this change of
coordinates the metric (\ref{ccd}) adopts a much more familiar expression:
\bea
&& \d s^2= 
 \frac{1}{\ff
(\la_1 r )\cos(\la_1 \rr) } \left( \d \rr^2  +\la_2^2\,\frac{\sin^2(\la_1 \rr)}{\la_1^2} \left( 
\d
 \te^2 + \frac{\sin^2(\la_2 \te)}{\la_2^2} \,\d\tes^2  \right) \right)  \nonumber\\
&&\quad\ \ =  \frac{1}{ \ff
(\la_1 r ) \cos(\la_1 \rr)}\, \d s_0^2 .
 \label{xd}
\eea
 Thus we have obtained a family of metrics,
parametrized by $\la_1,\la_2$ and depending on   the function     $\ff(\la_1 \rr) \equiv f(z\>q^2)$, which  is just
the  metric  $\d s_0$ of  the spaces of constant curvature~\cite{kiev}   multiplied by a global conformal
factor $  {1}/\left( \ff (\la_1 r ) \cos(\la_1 \rr)\right)$.  Hence in this approach the deformation parameter $\la_1$ governs the curvature and $\la_2$ provides the signature of the corresponding space.

Therefore, according to the pair $(\la_1,\la_2)$ (each non-zero $\la_i$ can be scaled to $\pm 1$ or $\pm {\rm i}$) we obtain  ``deformed analogues"~\cite{ober} of the 3D spherical   $(1,1)$,
hyperbolic   $({\rm i},1)$,  anti-de Sitter $(1,{\rm i})$ and de Sitter  $(\rm{i},\rm{i})$
spaces   with  {\em   non-constant} sectional    curvature. By one hand, these reduce
 to the proper flat  Euclidean $(0,1)$ and Minkowskian $(0,{\rm i})$ spaces under the contraction $\la_1\to 0$  (i.e.\  $z\to 0$):
$$
\d s^2\equiv   \d s_0^2=  \d \rr^2  +\la_2^2\,r^2 \left( 
\d
 \te^2 + \frac{\sin^2(\la_2 \te)}{\la_2^2} \,\d\tes^2  \right) .
  $$
On the other hand, the contraction $\la_2\to0$   leads to ``deformed"   oscillating and
expanding Newton--Hooke (NH)  spacetimes $(\la_1= 1,{\rm i})$, and whose limit $z\to 0$   gives  the   flat   Galilean
 spacetime with degenerate metric $\d s^2\equiv   \d s_0^2=  \d \rr^2$.   Nevertheless, in the following we shall avoid the Newtonian cases with $\la_2=0$ since their metric is
degenerate and their {\em direct} relationship with a 3D Hamiltonian is lost.
The specific  metrics of  these ``deformed" spaces are displayed in table~\ref{table1}, whereas the   connection and all the curvature tensors  for the generic metric  (\ref{xd})   are presented in table~\ref{table2}.

Now, the resulting
superintegrable {\em geodesic flow} Hamiltonian (\ref{ahaa}) on the above  family of curved spaces together with its {\em three}
constants of motion in the spherical-type  phase space  variables  turn out to be
\bea
&& {H}=\frac 12\, \ff(\la_1 r ) {\cos(\la_1 \rr)}
 \left( p_\rr^2  + \frac{\la_1^2}{\la_2^2\sin^2(\la_1 \rr)} \left(   
 p_\te^2 + \frac{\la_2^2}{\sin^2(\la_2 \te)}\,  p_\tes^2  \right)
\right) , \nonumber\\
&& 
{C}^{(2)}=p_\tes^2,\qquad 
{C}_{(2)}=\left(\cos\tes\, p_\te-\la_2\frac{\sin\tes\, p_\tes}{\tan(\la_2\te)}  \right)^2
\!\!,
\quad {C}^{(3)}= p_\te^2+
\frac{\la_2^2\,  p_\tes^2}{\sin^2(\la_2 \te)},
\label{mb}
\eea
where ${H}= 2{\cal H}$, ${C}^{(2)}=4{\cal C}^{(2)}$, ${C}_{(2)}= 4 \la_2^2 {\cal
C}_{(2)}$,  ${C}^{(3)}= 4 \la_2^2 {\cal C}^{(3)}$ and   $(p_\rr,p_\te,p_\tes)$ are the conjugate momenta of $(\rr,\te,\tes)$.
It is worth stressing that in this latter form the constants of motion $\{{C}^{(2)},{C}^{(3)},H\}$ directly induce the separability of the system:
\bea
&& {C}^{(2)}(\tes, p_\tes)=p_\tes^2 ,\quad {C}^{(3)}(\te,p_\te)= p_\te^2+
\frac{\la_2^2\,  {C}^{(2)}}{\sin^2(\la_2 \te)}, \nonumber \\
&& H(\rr,p_\rr)=\frac 12\, \ff(\la_1 r ) {\cos(\la_1 \rr)}
 \left( p_\rr^2  + \frac{\la_1^2 {C}^{(3)}}{\la_2^2\sin^2(\la_1 \rr)}  
\right) .
 \label{mh}
\eea

Notice that the expressions for the two specific examples previously commented, that is, the non-constant curved case with $\ff\equiv 1$  and the well-known spaces  of constant  curvature with $\ff=1/{\cos(\la_1 \rr)}$   studied in~\cite{ober} are directly recovered from these expressions.

%%%%%%%%%%%%%%%%%%%% table1 %%%%%%%%%%%%%%%%%%%%%%
\begin{table}[t] {\footnotesize
 \noindent
\caption{{Metric of  six 3D   spaces  of
non-constant curvature expressed in   spherical-type  coordinates  $(r,\te,\phi)$. The values of the parameters are 
 $z=\la_1^2\in\{\pm 1\}$ and $\la^2_2\in\{\pm 1,0\}$.  }}
\label{table1}
\medskip
\noindent\hfill
$$
\begin{array}{ll}
\hline
\\ [-6pt]
\mbox {$\bullet$ Deformed   spherical space ${\bf S}^3_z$} &\quad\mbox {$\bullet$ Deformed hyperbolic    space ${\bf
H}^3_z$ } \\[4pt]  z=+1;\ (\la_1,\la_2)=(1,1) &\quad z=-1;\  (\la_1,\la_2)=({\rm i},1)   \\[4pt]
\displaystyle{\d s^2 = \frac{1}{ \ff( r) \cos\rr} \left(  \d \rr^2 +  {\sin^2 \rr } \left( 
\d \te^2 +  {\sin^2  \te }  \,\d\tes^2 \right)  \right)  }  &\quad
\displaystyle{\d s^2 = \frac{1}{\ff({\rm i} r ) \cosh\rr} \left(  \d \rr^2    +\sinh^2  \rr   \left( 
\d \te^2 +  {\sin^2  \te } \,\d\tes^2 \right) \right) }   \\[12pt]
\mbox {$\bullet$ Deformed oscillating NH spacetime  ${\bf NH}^{2+1}_{+,z}$} &\quad\mbox {$\bullet$ Deformed
expanding NH spacetime  ${\bf NH}^{2+1}_{-,z}$ }\\[4pt] z=+1;\ (\la_1,\la_2)=(1,0) &\quad z=-1;\ 
(\la_1,\la_2)=({\rm i},0)   \\[4pt]
 \displaystyle{\d s^2 =  \frac{1}{\ff( r) \cos\rr }  \,   \d \rr^2   
 }  &\quad
 \displaystyle{\d s^2 =   \frac{1}{ \ff({\rm i} r ) \cosh\rr  }  \,  \d \rr^2    } \\[12pt]
\mbox {$\bullet$ Deformed anti-de Sitter  spacetime ${\bf AdS}^{2+1}_z$ } &\quad\mbox {$\bullet$ Deformed de
Sitter spacetime ${\bf dS}^{2+1}_z$}\\[4pt] z=+1;\ (\la_1,\la_2)=(1,{\rm i}) &\quad z=-1;\  (\la_1,\la_2)=({\rm
i},{\rm i})  \\[4pt]
\displaystyle{\d s^2 =  \frac{1}{\ff( r) \cos\rr } \left(   \d \rr^2  -{\sin^2  \rr }  \left( 
\d \te^2 +  \sinh^2  \te \,\d\tes^2 \right)  \right) }   &\quad
\displaystyle{\d s^2 =   \frac{1}{ \ff({\rm i} r )\cosh\rr } \left(  \d \rr^2 - \sinh^2 \rr  \left( 
\d
 \te^2 +  \sinh^2  \te  \,\d\tes^2 \right) \right) } \\[8pt]
\hline
\end{array}
$$
\hfill}
\end{table}
%%%%%%%%%%%%%%%%%%%%%%%%%%%%%%%%%%%%%%%%%%%%%%%%%%%%%%%%%%%%%%%%%%%%

%%%%%%%%%%%%%%%%%%%% table2 %%%%%%%%%%%%%%%%%%%%%%
\begin{table}[h]
{\footnotesize
 \noindent
\caption{{Non-zero components of the   connection, Riemann
 and Ricci   tensors, sectional  and scalar   curvatures of the metric (\ref{xd}) written in
 spherical-type coordinates $(r,\theta,\phi)$ for the  3D spaces with $sl_z(2)$
symmetry. We denote  $\ff^\prime=\frac{{\rm d}\ff(y)}{{\rm d}
 y}$ and $\ff^{\prime\prime}=\frac{{\rm d}^2\ff(y)}{{\rm d} y^2}$ with  
$y=\la_1 r$.
}}
\label{table2}
\medskip
\noindent\hfill
$$
\begin{array}{l}
\hline
\\[-6pt]
\mbox {$\bullet$ Connection} \\[4pt]
\displaystyle{
\Gamma^r_{r r}= \frac{\la_1}2\left(  \tan(\la_1
r)-\frac{\ff^\prime}{\ff}\right)
\qquad
\Gamma^\te_{\tes
\tes}=-\frac{\sin(2\la_2\te)}{2\la_2} 
\qquad
\Gamma^\tes_{\tes \te}=\frac{\la_2}{\tan(\la_2\te)}
}\\[12pt]
\displaystyle{
 \Gamma^\te_{\te r}=\Gamma^\tes_{\tes
r}={\la_1} \left( \frac{1+\cos^2(\la_1 r)}{\sin(2\la_1
r) }-
\frac{\ff^\prime}{2\ff} \right)\qquad \Gamma^r_{\te \te}=
-\frac{\la_2^2 \sin^2(\la_1 r)}{\la_1^2}\, \Gamma^\te_{\te r}
}\\[14pt]
\displaystyle{
 \Gamma^r_{\tes \tes}=- \frac{ \sin^2(\la_1 r)}{\la_1^2} \, 
\sin^2(\la_2\te)\Gamma^\te_{\te r}
}\\[12pt]
\mbox {$\bullet$ Riemann tensor  }\\[2pt]
\displaystyle{
R^\te_{r\te r}=R^\tes_{r\tes r}=\frac 12 \la_1^2\left(\frac{1}{\tan(\la_1
r)}\,\frac{\ff^\prime}{\ff}+\frac{ \ff^{\prime\prime}}{\ff} -\left( \frac{\ff^\prime}{\ff}
\right)^2-\tan^2(\la_1 r) \right)  }\\[12pt]
\displaystyle{ R^r_{\te r\te}=\la_2^2\,\frac{\sin^2(\la_1 r)}{\la_1^2}\,R^\te_{r\te r}
\qquad  R^r_{\tes r\tes}= \frac{\sin^2(\la_1 r)}{\la_1^2}\,\sin^2(\la_2 \te)\,R^\te_{r\te r}
}\\[12pt]
\displaystyle{
R^\tes_{\te\tes \te}=   \la_2^2\left(1-\sin^2(\la_1 r)\left(
\frac{1+\cos^2(\la_1 r)}{\sin(2\la_1 r) }-
\frac{\ff^\prime}{2\ff}\right)^2\,\right)\qquad
R^\te_{\tes\te \tes}=\frac{\sin^2(\la_2 \te)}{\la_2^2}  R^\tes_{\te\tes \te} 
}\\[14pt]
\mbox {$\bullet$ Ricci tensor  }\\[2pt]
\displaystyle{
R_{r r}=  \la_1^2\left(\frac{1}{\tan(\la_1
r)}\,\frac{\ff^\prime}{\ff}+\frac{ \ff^{\prime\prime}}{\ff} -\left( \frac{\ff^\prime}{\ff}
\right)^2-\tan^2(\la_1 r) \right) 
\qquad
R_{\tes \tes}=\frac{\sin^2(\la_2 \te)}{\la_2^2}  R_{\te \te}
}\\[12pt]
\displaystyle{
R_{\te \te}=  \la_2^2 \sin^2(\la_1 r)  \left(
 \frac{(1+2\cos^2(\la_1 r))}{\sin(2\la_1 r) }\,\frac{\ff^\prime}{\ff}
+\frac{ \ff^{\prime\prime}}{2\ff} -\frac 34 \left(
\frac{\ff^\prime}{\ff}
\right)^2-\frac{3}{4}\tan^2(\la_1 r)\right) 
}\\[14pt]
\mbox {$\bullet$ Sectional curvatures  }\\[2pt]
\displaystyle{
K_{r\te}=K_{r\tes}= \frac 12\, \la_1^2\cos(\la_1 r) 
\left( \frac{1}{\tan(\la_1 r)}\, \ff^\prime + \ff^{\prime\prime} -
\frac{ {\ff^\prime}^2} {\ff}
 -\tan^2(\la_1 r) \ff\right)  
}\\[12pt]
\displaystyle{
K_{\te\tes}=   \la_1^2 \cos(\la_1 r)\, \ff  \left(\frac{1}{\sin^2(\la_1 r)}-\left(
\frac{1+\cos^2(\la_1 r)}{\sin(2\la_1 r) }-
\frac{\ff^\prime}{2\ff}\right)^2\,\right) 
}\\[14pt]
\mbox {$\bullet$ Scalar curvature}\\[2pt]
\displaystyle{
K = 2 \la_1^2\cos(\la_1 r) 
\left(  \frac{(1+3\cos^2(\la_1 r))}{\sin(2\la_1 r) } \, \ff^\prime +  \ff^{\prime\prime} -
\frac 5 4\,\frac{ {\ff^\prime}^2} {\ff}
 -\frac 54\,\tan^2(\la_1 r) \ff\right)  
}\\[12pt]
\hline
\end{array}
$$
\hfill}
\end{table}
%%%%%%%%%%%%%%%%%%%%%%%%%%%%%%%%%%%%%%%%%%%%%%%%%%%%%%%%%%%%%%%%%%%%

%%%%%%%%%%%%%%%%%%%%%%%%%%%%%%%%%%%%%%%%%%%%%%%%%%%
\section{Superintegrable  potentials}

The results of proposition 1  allows to construct many types of  superintegrable
potentials  on 3D curved spaces through specific choices of the Hamiltonian function
${\cal H}$ (\ref{ham})  which could be
momenta-dependent potentials, central ones, centrifugal terms, etc.\ (see~\cite{jpa2D} for the 2D case).
Our aim now is, firstly,  to characterize in this $sl_z(2)$ coalgebra framework  the superposition of central potentials with (up to) three centrifugal terms and, secondly, to single out which would be the corresponding intrinsic KC and oscillator potentials. Hereafter we shall mainly make use of the spherical-type phase space introduced in the previous section since these variables   will allow us to deal with both curved Riemannian and Lorentzian spaces simultaneously.

%%%%%%%%%%%%%%%%%%%%%%%%%%%%%%

\subsection{Central potentials with centrifugal terms}

If we consider the 3D symplectic realization of $sl_z(2)$ (\ref{be}),  with arbitrary $b_i$'s,   and 
we add a   smooth function ${\cal V}(z \jm )$ to   the free Hamiltonian (\ref{ahaa}),    
\be
 {\cal H}=\frac 12 \jp\, f (z\jm )+{\cal V}(z \jm ),
\label{xfa}
\ee
 we obtain a system formed by the superposition of a central potential with three centrifugal $b_i$-terms. By taking into account that, in  terms of the spherical-type variables, ${\rm e}^{-z \jm}=\cos(\la_1\rr)$, ${\cal V}(z \>q^2 ) \equiv   {\cal U}( \la_1\rr)$  and 
 \bea
 && J_+= \cos(\la_1\rr) \left( p_\rr^2  + \frac{\la_1^2}{\la_2^2\sin^2(\la_1 \rr)} \left(   
 p_\te^2 + \frac{\la_2^2\,  p_\tes^2}{\sin^2(\la_2 \te)}  \right)  \right) \nonumber\\
 &&\qquad +\frac{\la_1^2\cos(\la_1\rr)}{\sin^2(\la_1\rr) } \left( \frac{b_1}{\cos^2(\la_2\te)} +   \frac{b_2\la_2^2 }{\sin^2(\la_2\te)\cos^2\tes}+  \frac{b_3\la_2^2 }{\sin^2(\la_2\te)\sin^2\tes}
  \right)  ,\nonumber
\eea
  the Hamiltonian (\ref{xfa})    turns out to be
\bea
&&\!\!\!\!\!\!\!\! \!\!\!\!\!\!\!\! 
 {H} =\frac 12\, \ff(\la_1 r ) \, {\cos(\la_1 \rr)}
 \left( p_\rr^2  + \frac{\la_1^2}{\la_2^2\sin^2(\la_1 \rr)} \left(   
 p_\te^2 + \frac{\la_2^2\,  p_\tes^2}{\sin^2(\la_2 \te)}  \right)  \right)  + {\cal U}( \la_1\rr) \nonumber\\
&&  +\frac{\la_1^2\,\ff(\la_1 r )\cos(\la_1\rr)}{2\sin^2(\la_1\rr) } \left( \frac{b_1}{\cos^2(\la_2\te)} +   \frac{b_2\la_2^2 }{\sin^2(\la_2\te)\cos^2\tes}+  \frac{b_3\la_2^2 }{\sin^2(\la_2\te)\sin^2\tes}
  \right) ,
\label{ffa}
\eea
which  is, by construction, {\em superintegrable}. We stress that, for any choice of $f$ and $\cal U$, they share the same set of {\em three} constants of the
motion  (\ref{bbg}), which now depend on $\la_1$ and $\la_2$;  these  are explicitly given, up to some additive and multiplicative constants,  by
\bea
&& 
{C}^{(2)}=p_\phi^2+\frac{b_2\la^2_2}{\cos^2 \phi}
+\frac{b_3\la^2_2}{\sin^2\phi}, \nonumber\\
&&
 {C}_{(2)}=\left(\cos\phi\, p_\theta-\frac{\la_2\sin\phi\, p_\phi}{\tan(\la_2\theta)} \right)^2+b_1\la_2^2\tan^2(\la_2\theta)\cos^2\phi
+\frac{b_2\la^4_2}{\tan^2(\la_2\theta)\cos^2\phi}, \nonumber\\
&&
{C}^{(3)}=p_\theta^2+\frac{\la_2^2 p_\phi^2 }{ 
\sin^2(\la_2\theta)}
+\frac{b_1\la_2^2}{\cos^2(\la_2\theta)} +\frac{b_2\la_2^4}{\sin^2(\la_2\theta)\cos^2\phi}
+\frac{b_3\la_2^4}{\sin^2(\la_2\theta)\sin^2\phi}  .
\label{ffb}
\eea
Hence the constants of motion
${C}^{(2)}$ and ${C}^{(3)}$ and the Hamiltonian can be written as:
\bea
&&  {{C}^{(2)}(\phi,p_\phi)=p_\phi^2+\frac{b_2\la^2_2}{\cos^2 \phi}
+\frac{b_3\la^2_2}{\sin^2\phi} ,}\nonumber \\
&& {{C}^{(3)}(\theta,p_\theta)=p_\theta^2+\frac{b_1\la_2^2}{\cos^2(\la_2\theta)} +   \frac{\la_2^2}{ 
\sin^2(\la_2\theta)}\,{C}^{(2)}
,}\nonumber \\
&& { {H}(\rr,p_\rr)=\frac 12  \ff(\la_1 r ) {\cos(\la_1 \rr)} 
 \left(   p_\rr^2+
\frac{\la_1^2 {C}^{(3)}   }{\la_2^2 \sin^2(\la_1 \rr) } \right)+ {\cal U}( \la_1\rr) .}
\label{zxa}
\eea
Therefore, similarly to the free motion, the Hamiltonian (\ref{ffa}) is separable and  reduced to a 1D radial system.  
 
%%%%%%%%%%%%%%%%%%%%%%%%%%%%%%%%%%%%%%%%%%%%%%%%%%%\

\subsection{Intrinsic Kepler--Coulomb  and oscillator  potentials}

In order to define the  ``intrinsic" KC and oscillator potentials on the 3D curved Riemannian and Lorentizan spaces with metric (\ref{xd}) given in table~\ref{table1} we shall apply the approach introduced in~\cite{LT87, LT95, EP06d, CQG} for 3D spherically symmetric spaces. This requires to transform the metric (\ref{xd})  by introducing 
$$
\Theta:=\la_2\theta,\quad R:=\la_1 r,\quad h(R):=\cos(\la_1 r)g(\la_1
r) ,
\label{fh}
$$
which yields
$$
\dd s^2=\frac1{\la_1^2 h(R)}\left( \dd
R^2+\sin^2R\left(\dd\Theta^2+\sin^2\Theta\,\dd\phi^2 \right)\right) .
\label{fg}
$$
Next the radial symmetric Green function $U(R)$ on the curved space (up to multiplicative and additive
constants) is defined as the positive non-constant solution to the equation
$$
\Delta_{\rm LB} U(R)=\frac{\la_1^2h(R)^{3/2}}{\sin^2R}\frac{\dd}{\dd
R} \left(\frac{\sin^2R}{\sqrt{h(R)}}\frac{\dd U(R)}{\dd R}\right)=0,
$$
where $\Delta_{\rm LB}$ is the  (intrinsic) Laplace--Beltrami operator in the above
coordinates. Then
\be
U(\la_1 r)\equiv U(R)=\int^R \frac{\sqrt{h(R')}}{\sin^2R'}\dd R'.
\label{fi}
\ee
And the intrinsic KC and oscillator potentials are defined by
\begin{equation}
{\cal U}\Kep(\la_1 r) :=\cte\,U (\la_1 r) ,\qquad {\cal U}\Harm(\la_1 r) :=\frac \cteb{U^2(\la_1 r)},
\label{wd}
\end{equation}
 where  $\cte$ and $\cteb$ are     real constants.

In what follows we illustrate these results through some examples according to some particular choices of the function $g$.

\subsubsection{Constant curvature: $g(\la_1 r)=1/\cos(\la_1 r)= \exp({z J_-})$}

   To start with let us consider the very singular case of {\em constant curvature} with   $h(R)=1$ so that the Green function (\ref{fi})  turns to be
$$
U(R)=-\frac{1}{\tan R}.
$$
Consequently we recover the well-known KC and oscillator potentials~\cite{kiev, Higgs, RS, Schrodingerdual,18,Schrodingerdualb,car1,Sh05} for the spherical and (anti-)de Sitter spaces with $\la_1$ real, and for the hyperbolic and de Sitter spaces with $\la_1$ imaginary:
$$
{\cal U}\Kep(\la_1 r)=-\frac{\cte\la_1}{\tan(\la_1 r)},\quad  {\cal U}\Harm(\la_1 r)= \cteb \,\frac{\tan^2 (\la_1 r) }{\la_1^2}.
$$
The scalar curvature is $K=6 \la_1^2$.  Obviously, the limit $\la_1\to 0$ is well defined and leads to flat Euclidean/Minskowskian potentials: 
 $-\cte/r$ and $\cteb r^2$. In terms of the $sl_z(2)$ generators the resulting Hamiltonians are expressed by
$$
{\cal H} \Kep =\frac 12 \, J_+ {\rm e}^{z J_-} - \cte \sqrt{  \frac{z \,{\rm e}^{-z \jm}  }{ \sinh(z J_-)}} ,
$$
$$
{\cal H} \Harm =\frac 12 \, J_+ {\rm e}^{z J_-}+ \cteb \,    \frac{ \sinh(z J_-) }{z} \,{\rm e}^{ z \jm} ,
$$
which were already given in~\cite{jpa2D} for the 2D case. Recall that the three centrifugal terms are directly obtained from $J_+$ by considering a symplectic realization with non-vanishing $b_i$'s. In this way the generalized KC Hamiltonian~\cite{Evansa, kiev,Miguel, Williamsx,vulpiPAN} and the curved Smorodinsky--Winternitz system~\cite{kiev, RS,18, vulpiPAN,FMSUW65, Ev90b, 8, 10, 11, 20, 21,BHSS03} are recovered.

\subsubsection{$g(\la_1 r)=1$}

The case $g\equiv 1$, which was commented in the previous sections, leads to  $h(R)=\cos R$. Now the Green function involves an elliptic integral of the second kind, $E( x | m)$:
 $$
 U(\la_1 r) =-\frac{\la_1}{\tan(\la_1 r)}\, \sqrt{\cos(\la_1 r)} - \la_1 E\biggl(\frac 12\, \la_1 r \, | \, 2\biggr) .
 $$
 Note that a multiplicative constant $\la_1$ has been introduced in order to ensure a well defined non-deformed limit (flat contraction):
 $\lim_{\la_1\to 0} U(\la_1 r)=-1/r$.

This can be considered as  the ``simplest" case from the $sl_z(2)$ viewpoint, since the kinetic term is just  $\frac 12 J_+$. In spite of this fact,  the scalar curvature (which can be obtained from the general expression given in table~\ref{table2}) reads
$$
K= -\frac 52\, \la_1^2 \sin(\la_1 r) \tan(\la_1 r)  ,
$$
and the associated KC and oscillator potentials are otained from (\ref{wd}). In this respect, we remark that these potentials turn out to be much more complicated than the ones coming from the Ansatz proposed in~\cite{jpa2D}.

\subsubsection{$g(\la_1 r)=(\cos(\la_1 r))^{4 k-1}= \exp({-  z (4k -1)J_-})$ for a real constant $k\ne 1$}

 The two aforementioned examples can be included in this more general class of systems as they are reproduced for $k=0$ and $k=1/4$, respectively.  Starting from $h(R)=(\cos R)^{4 k}$ it is found that the resulting Green function 
depend on a hypergeometric function ${}_2 F_1(a,b,c,x)$ that can also be expressed by means of an incomplete beta function $B(x,a,b)$ in the form
 \bea
&& U(\la_1 r)=\frac{(\sin(\la_1 r))^{2(k-1)}  }{2(k-1) }  \  {}_2 F_1\left(1-k,\frac 12 -k ,2-k,\frac{1}{\sin^2(\la_1 r)}\right) \nonumber \\
&&\qquad \quad\ =- \frac 12 B\left(\frac{1}{\sin^2(\la_1 r)} , 1-k,\frac 12 +k\right) \nonumber
\eea
  provided that we have dropped  a multiplicative constant ${\rm i} (-1)^{-k}$.
 The scalar curvature of the underlying spaces is
 $$
K= \la_1^2 (\cos(\la_1 r))^{4 k-2} \biggr (3- 4 k (k+4) + (3+ 4 k (k-2) \cos(2\la_1 r)  \biggl)   .
$$

 \subsubsection{$g(\la_1 r)=\cos^3(\la_1 r)= \exp({- 3 z J_-})$}

To end with, we study the forbidden value  $k=1$ in the previous example, which corresponds to take 
$h(R)=\cos^4 R$. Then we obtain the following closed expression for the Green function
$$
U(\la_1 r)=-\la_1^2 r - \frac{\la_1}{\tan(\la_1 r)} ,
$$
where we have introduced a multiplicative constant $\la_1$ such that $\lim_{\la_1\to 0}U(\la_1 r)=-1/r$.
This yields the corresponding   KC and oscillator potentials 
$$
{\cal U}\Kep(\la_1 r)=-\cte \left( \la_1^2 r+\frac{\la_1}{\tan(\la_1 r)}\right),\quad
   {\cal U}\Harm(\la_1 r)= \cteb \,\frac{\tan^2 (\la_1 r) }{\la_1^2(1+\la_1 r\tan(\la_1 r)^2},
$$
which are worth to be compared with the corresponding expressions of the constant curvature case.
Finally, the scalar curvature for this space is:
$$
K=-\la_1^2 \cos^2(\la_1 r) (17+\cos(2\la_1 r)) .
$$

%%%%%%%%%%%%%%%%%%%%%%%%%%%%%%%%%%%%%%%%%%%%%%%%%%%%%%%%%%%%%%%%%%%%%%%%%%%%%%%%%

\section*{Acknowledgements}

This work was partially supported by the Spanish Ministerio de Educaci\'on   under grant no.\    MTM2007-67389 (with EU-FEDER support)  (A.B.\ and F.J.H.), by the Spanish DGI and CAM--Complutense University under grants  no.~FIS2008-00209 and~CCG07-2779 (A.E.), and by the INFN--CICyT (O.R.). O.R. thanks the Einstein Foundation and Russian Foundation for
Basic Research for supporting the research project  ``Integrable/solvable
Classical and Quantal Many-Body Problems and their integrable
discretizations".

 \newpage

{\footnotesize

\end{document}